\documentclass[english]{sbrt}
\usepackage[english]{babel}
\usepackage[utf8]{inputenc}
\usepackage{graphicx}

\usepackage{stfloats}
\usepackage{graphicx}
\usepackage{multirow}
\usepackage{xcolor}

\usepackage{amsmath,amsmath,amsfonts,upgreek}
\usepackage{balance}

\begin{document}

\title{CVNN-based Channel Estimation and Equalization \\ in OFDM Systems Without Cyclic Prefix}

\author{Heitor dos Santos Sousa, Jonathan Aguiar Soares, Kayol Soares Mayer, and Dalton Soares Arantes
\thanks{Heitor S. Sousa, Jonathan A. Soares, Kayol S. Mayer, and Dalton S. Arantes are with the Digital Communications Laboratory -- ComLab, Department of Communications, School of Electrical and Computer Engineering, University of
Campinas -- Unicamp, 13083-852 Campinas, SP, Brazil, e-mails: h208981@dac.unicamp.br; j229966@dac.unicamp.br; kayol@unicamp.br; dalton@unicamp.br.}
}

\maketitle

\markboth{XLI SIMPÓSIO BRASILEIRO DE TELECOMUNICA\c{C}\~{O}ES E PROCESSAMENTO DE SINAIS - SBrT 2023, 08--11 DE OUTUBRO DE 2023, SÃO JOSÉ DOS CAMPOS, SP}{}

\begin{abstract}

In modern communication systems operating with Orthogonal Frequency-Division Multiplexing (OFDM), channel estimation requires minimal complexity with one-tap equalizers. However, this depends on cyclic prefixes, which must be sufficiently large to cover the channel impulse response. Conversely, the use of cyclic prefix (CP) decreases the useful information that can be conveyed in an OFDM frame, thereby degrading the spectral efficiency of the system. In this context, we study the impact of CPs on channel estimation with complex-valued neural networks (CVNNs). We show that the phase-transmittance radial basis function neural network offers superior results, in terms of required energy per bit, compared to classical minimum mean-squared error and least squares algorithms in scenarios without CP.
\end{abstract}

\section{Introduction}

With the increasing demand for technologies such as massive machine-type communications~(mMTC), enhanced mobile broadband~(eMBB), and ultra-reliable and low latency communications~(URLLC), higher data rates over tighter bandwidths have never been so important~\cite{Hsu2023}. In view of these demands, recent telecommunications technologies (e.g., 5G and beyond) have become paramount to support these services.

A key enabling technology in most current communication systems is orthogonal frequency-division multiplexing~(OFDM). Employing OFDM, communications systems can convey information via multiple and closely spaced sub-carriers (i.e., narrowband subchannel frequencies). Besides the natural user multiplexing in different frequencies, OFDM also provides advantages regarding intersymbol interference~(ISI)~\cite{Xia2023} and intercarrier interference~(ICI)~\cite{Liao2023}. While the former is simply avoided by using a cyclic prefix, the latter is only a problem in dynamic channels or local oscillator mismatches with high carrier frequency offsets~(CFOs)~\cite{Cheng2021}.

In typical OFDM systems, to increase the useful data rate (not considering channel coding), it is possible to reduce the pilot rate, extend the number of subcarriers, or increase the modulation order. However, these solutions come with their own set of issues. By reducing the pilot rate, the channel estimation becomes less accurate, and the system becomes more susceptible to rapid channel variations, such as those present in dynamic channels. By extending the number of subcarriers while maintaining the same bandwidth (to avoid interference in adjacent channels), there is an increase in both the computational complexity and the clock speed required to process the transmitted and received signals. Lastly, increasing the modulation order degrades the bit error rate (BER) at the receiver.


Motivated by the widespread use of machine learning~(ML) and bolstered by the support of the universal approximation theorem of artificial neural networks~(ANNs)~\cite{Hornik1989}, several works have proposed ML-based algorithms for channel estimation in OFDM systems~\cite{Mei2021a, Soares2022, Le2021, Mei2021b, Jebur2021, Muller2023, Ye2018}. Soares et al.~\cite{Soares2022} extended a classical smoothing filter of channel estimation with principal component analysis~(PCA) to cut residual noise. Le et al.~\cite{Le2021} improve the least squares~(LS) channel estimation results using fully connected deep neural networks (DNNs), convolutional neural networks~(CNNs), and bidirectional long short-term memory~(LSTM). Mei et al.~\cite{Mei2021b} employed online and linear ML-based channel estimation with low complexity and a fast convergence rate. Jebur et al.~\cite{Jebur2021} implemented an ML-based channel estimator to track time-varying and frequency-selective channels with a small number of pilots. M\"{u}ller et al.~\cite{Muller2023} proposed a symbol timing recovery algorithm with deep radial basis function~(RBF) ANNs. Ye et al.~\cite{Ye2018} demonstrated the potential of deep learning for joint channel estimation and signal detection of OFDM systems without cyclic prefix~(CP). 

Apart from these real-valued ML algorithms, complex-valued neural networks~(CVNNs) have also presented promising results for telecommunications, such as channel equalization, beamforming, channel estimation, and decoding~\cite{Liu2019, Mayer2019, Enriconi2020, Mayer2020, Soares2021,Mayer2022,MayerThesis}. As already demonstrated in the literature, compared with real-valued neural networks~(RVNNs), CVNNs have increased functionality, better performance, and reduced training time~\cite{Hirose2012,Cruz2022, Zhang2022}. Furthermore, CVNNs also rely on the universal approximation theorem, as recently proved in~\cite{Voigtlaender2023}.

In this context, this paper proposes an extension of~\cite{Ye2018} for OFDM channel estimation and equalization using a CVNN, the phase-transmittance radial basis function~(PT-RBF) neural network, in a CP-free scenario to increase the useful information rate. Unlike~\cite{Ye2018}, in this work we do not take into account the OFDM decoding since it can be easily handled with a low complexity fast Fourier transform~(FFT). Moreover, it reduces the search space, and, consequently, the neural network complexity, i.e., a smaller number of layers and neurons is necessary. This paper aims to demonstrate a suitable approach to online channel estimation of OFDM without CP. It is important to note that, similarly to this paper, Chu et al.~\cite{Chu2022} proposed a channel estimation technique using a CVNN for optical systems operating with filter bank multicarrier with offset quadrature amplitude modulation~(FBMC/OQAM). However, compared to OFDM, FBMC has a higher implementation complexity, and the absence of a cyclic prefix~(CP) is an intrinsic characteristic of this architecture. To the best of our knowledge, it is the first work handling a CVNN for OFDM channel estimation and equalization without CP.


The remainder of this paper is organized as follows. Section~\ref{sec:OFDM} discusses the implemented OFDM communication scheme, and Section~\ref{sec:channel_estimation} the classical least squares~(LS) and minimum mean-square error~(MMSE) channel estimation algorithms. Section~\ref{sec:ptrbf} describes the proposed PT-RBF channel estimation. Section~\ref{sec:comp_complex} presents the computational complexities of the proposed algorithm. Section~\ref{sec:results} presents the results of the proposed approach compared with LS and MMSE channel estimation. Lastly, Section~\ref{sec:conc} concludes the paper.

\section{OFDM System Model}
\label{sec:OFDM}

\begin{figure}[t]
\centering\includegraphics[width=\columnwidth]{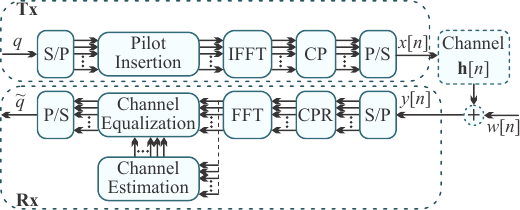}
    \caption{Orthogonal frequency division multiplexing~(OFDM) communication architecture.}
    \label{fig:setup}
\end{figure}

Fig.~\ref{fig:setup} illustrates the OFDM communication architecture considered in this work. First, on the transmitter~(Tx) side, the QAM symbols $q$ are parallelized by an S/P block, mapping $\mathbb{C} \mapsto \mathbb{C}^{K}$, where $K$ is the number of sub-carriers. Then, the $K$-parallel symbols feed the pilot insertion block that, depending on the channel estimation scheme at the receiver~(Rx) side, inserts block- or comb-type pilots~\cite{Coleri2002}. Next, an inverse fast Fourier transform~(IFFT) block converts data from the frequency domain to the time domain. In the sequel, a cyclic prefix~(CP) of length $N_{cp}$ is inserted at the beginning of each IFFT output to mitigate the inter-symbol interference~(ISI)~\cite{Ye2018}. Finally, the resultant signal is serialized~(i.e., $\mathbb{C}^{K+N_{cp}} \mapsto \mathbb{C}$) in the parallel to serial~(P/S) block and sent over the wireless channel.

Considering a sample-spaced multipath channel with $N_{ds}$ samples $\mathbf{h}[n]\in\mathbb{C}^{N_{ds}}$, the baseband received signal is
\begin{equation}
\label{eq:channel}
y[n] = \sum_{i=0}^{N_{ds}-1}h_{i+1}[n]x[n-i]+w[n], 
\end{equation}
where $n$ is the discrete-time index, $x[n]\in\mathbb{C}$ is the transmitted data, and $w[n]\sim \mathcal{C}\mathcal{N}(0,\,\sigma_w^{2})\in\mathbb{C}$ is the additive white Gaussian noise~(AWGN) at the receiver, with zero mean and variance~$\sigma_w^{2}$. 

On the receiver side, the income signal is firstly parallelized by an S/P block, mapping $\mathbb{C} \mapsto \mathbb{C}^{K+N_{cp}}$. Next, the cyclic prefix is removed in the CPR block, and the resultant signal of dimension $\mathbb{C}^K$ is converted to the frequency domain by a fast Fourier transform~(FFT) block. Afterward, the FFT output feeds the channel estimation block, which estimates the channel for the equalization block. A P/S block serializes the resulting signal to produce the estimated outputs $\hat{q}$.

\section{Channel Estimation and Equalization}
\label{sec:channel_estimation}

In this work, we consider block-type pilot arrangements for channel estimation. In the frequency domain, we consider $x[k]$ the pilots before the CP block at the Tx, and $y[k]$ the pilots after the FFT block at the Rx, with the frequency domain index $k \in [1, 2, \cdots, K]$. In addition, $h[k]$ corresponds to the channel related to the $k$-th subcarrier. 

The LS estimator minimizes the square distance between $x[k]$ and $y[k]$~\cite{Savaux2017}, obtaining the channel estimation
\begin{equation}
\label{eq:ls}
    h_{LS}[k] = \frac{y[k]}{x[k]}.
\end{equation}

The MMSE estimator is obtained with second-order statistics of the channel~\cite{Savaux2017}, to minimize the mean square error, as
\begin{equation}
\label{eq:mmse}
    h_{MMSE}[k] = \sigma^2_{h}[k]\left(\sigma^2_{h}[k]|x[k]|^2+\sigma^2_{w}[k] \right)^{-1}x^*[k]y[k],
\end{equation}
in which $[\cdot]^{*}$ is the conjugate operator, $[\cdot]^{-1}$ is the inverse operator, and $\sigma^2_{h}[k]$ and $\sigma^2_{w}[k]$ are the variances of $h[k]$ and $w[k]$, respectively. Note that, besides the inherent higher computational complexity compared with \eqref{eq:ls}, the MMSE estimator is also dependent on the channel statistics $\sigma^2_{h}[k]$.

In both LS and MMSE estimations, the equalization is performed, per subcarrier, as
\begin{equation}
    \tilde{x}[k] = \frac{y[k]}{\tilde{h}[k]},
\end{equation}
where $\tilde{h}[k]=h_{LS}[k]$ or $\tilde{h}[k]=h_{MMSE}[k]$, depending on the channel estimation method.

\section{Proposed CVNN Channel Estimation and Equalization}
\label{sec:ptrbf}

\begin{figure}[t]
\centering
  \includegraphics[width=\columnwidth]{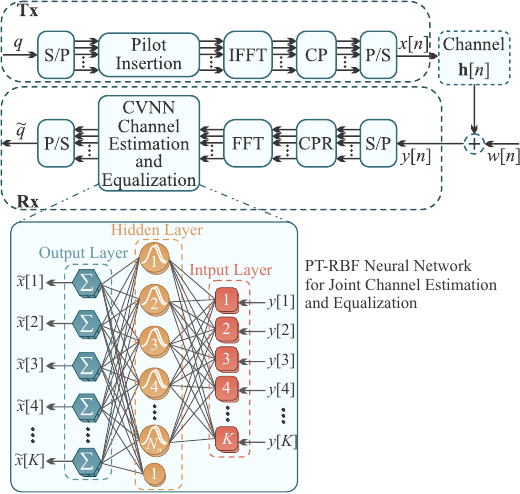}
    \caption{Orthogonal frequency division multiplexing~(OFDM) communication architecture with the proposed CVNN-based channel estimation and equalization. The CVNN block represents the PT-RBF neural network composed of three layers.}
    \label{fig:proposed}
\end{figure}

The proposed channel estimation and equalization with a CVNN are shown in Fig.~\ref{fig:proposed}. Unlike the usual channel estimation and equalization illustrated in Fig.~\ref{fig:setup}, the proposed ML scheme performs both tasks at the same time, in a joint scheme. The CVNN block is composed of a PT-RBF neural network with three layers. The input layer is fed with the FFT output vector $\mathbf{y}$. The hidden layer maps $\mathbf{y}$ onto a nonlinear space, through $N_n$ neurons. The Gaussian kernel (i.e., activation function) of the $i$-th neuron is
\begin{equation}
    \phi_i = \exp{\left[ \Re\{v_i\}\right]}+\jmath\exp{\left[ \Im\{v_i\}\right]},
\end{equation}
in which $v_i$ is the argument of the $i$-th Gaussian kernel, given by
\begin{equation}
    v_i = \frac{\Vert \Re\{\mathbf{y}\}-\Re\{\boldsymbol{\upgamma}_i\}\Vert^2_2}{\Re\{\sigma^2_i\}} +\jmath \frac{\Vert \Im\{\mathbf{y}\}-\Im\{\boldsymbol{\upgamma}_i\}\Vert^2_2}{\Im\{\sigma^2_i\}},
\end{equation}
where $\boldsymbol{\upgamma}_i\in\mathbb{C}^K$ is the center vector and $\sigma^2_i\in\mathbb{C}$ is the variance of the $i$-th neuron. The operators $\Re\{\cdot\}$ and $\Im\{\cdot\}$ return the real and imaginary components, respectively. The Euclidean norm is denoted by $\Vert \cdot \Vert_2$.

In the output layer, the $k$-th equalized output data $\tilde{x}[k]$ is obtained via a linear combination of the Gaussian kernels and the vector of synaptic weights $\mathbf{w}_k\in\mathbb{C}^{N_n}$, as 
\begin{equation}
    \tilde{x}[k]=\mathbf{w}_k^T\boldsymbol{\upphi},
\end{equation}
in which $[\cdot]^T$ is the transpose operator and $\boldsymbol{\upphi}\in\mathbb{C}^{N_n}$ is the vector of Gaussian kernels.

The PT-RBF is optimized via the complex-valued backpropagation process, using the stochastic gradient descent~(SGD)~\cite{MayerThesis}. For the sake of representation, $m$ denotes the SGD iteration during a training epoch with $M$ attributes, i.e., $M$-OFDM symbols from the preamble. As the PT-RBF neural network utilized in this work has only one hidden layer, its update equations can be simplified from the generalized model proposed in \cite{Mayer2022}, as
\begin{equation}
\label{eq:PTRBFupdateMatrices_weights}
    \mathbf{W}[m+1] = \mathbf{W}[m] + \eta_w \mathbf{e}[m]\boldsymbol{\upphi}^H[m], 
\end{equation}
\begin{equation}
\label{eq:PTRBFupdateMatrices_bias}
    \mathbf{b}[m+1] = \mathbf{b}[m] + \eta_b\boldsymbol{\upphi}[m], 
\end{equation}
\begin{multline}
\label{eq:PTRBFupdateMatrices_gamma}
    \boldsymbol{\Gamma}[m+1] =\boldsymbol{\Gamma}[m]+\eta_\gamma \Re\{\boldsymbol{\updelta}[m]\}\left[ \Re\{\mathbf{Y}_m\}-\Re\{\boldsymbol{\Gamma}[m]\} \right]\\
    +\jmath\eta_\gamma \Im\{\boldsymbol{\updelta}[m]\}\left[ \Im\{\mathbf{Y}_m\}-\Im\{\boldsymbol{\Gamma}[m]\} \right], 
\end{multline}
\begin{multline}
\label{eq:PTRBFupdateMatrices_sigma}
    \boldsymbol{\upsigma}^2[m+1] = \boldsymbol{\upsigma}^2[m]+\eta_\sigma\Re\{\boldsymbol{\updelta}[m]\}\Re\{\mathbf{v}[m]\}\\ +\jmath \eta_\sigma \Im\{\boldsymbol{\updelta}[m]\}\Im\{\mathbf{v}[m]\}, 
\end{multline}
where $\mathbf{v}\in\mathbb{C}^{N_n}$ is the vector of Gaussian kernel arguments, $\mathbf{W}\in\mathbb{C}^{K\times N_n}$ is the matrix of synaptic weights, $\mathbf{b}\in\mathbb{C}^K$ is the vector of bias, $\boldsymbol{\Gamma}\in\mathbb{C}^{N_n\times K}$ is the matrix of center vectors, and $\boldsymbol{\upsigma}^2\in\mathbb{C}^{N_n}$ is the vector of center variances. The parameters $\eta_w$, $\eta_b$, $\eta_\gamma$, and $\eta_\sigma$ correspond to the learning rate of $\mathbf{W}$, $\mathbf{b}$, $\boldsymbol{\Gamma}$, and $\boldsymbol{\upsigma}^2$, respectively. The operator $[\cdot]^H$ denotes the Hermitian.

For $\boldsymbol{\Gamma}$ and $\boldsymbol{\upsigma}^2$, the diagonal matrix of local gradients $\boldsymbol{\updelta}\in\mathbb{C}^{N_n\times N_n}$ is given as 
\begin{multline}
\label{eq:local_grad}
    \boldsymbol{\updelta}[m] = \mathrm{diag}[\Re\{\mathbf{W}^H[m] \mathbf{e}[m]\}\odot\Re\{\boldsymbol{\upphi}[m]\}\oslash\Re\{\boldsymbol{\upsigma}^2[m]\}]\\
    +\jmath\mathrm{diag}[\Im\{\mathbf{W}^H[m] \mathbf{e}[m]\}\odot\Im\{\boldsymbol{\upphi}[m]\}\oslash\Im\{\boldsymbol{\upsigma}^2[m]\}], 
\end{multline}
in which $\mathrm{diag}[\cdot]$ returns a square diagonal matrix with its argument vector on the main diagonal, and $\odot$ and $\oslash$ denote the Hadamard product and division, respectively. The error vector is given by $\mathbf{e}[m]=\mathbf{x}_m-\mathbf{\tilde{x}}_m$, for the PT-RBF output $\mathbf{\tilde{x}}_m$ with respect to the desired OFDM symbol $\mathbf{x}_m$, in the preamble. In addition, $\mathbf{Y}_m$ is the expanded matrix of OFDM input symbols
\begin{equation} 
\label{eq:expanded_matrix}
    \mathbf{Y}_m=\begin{bmatrix}
                    \textbf{\text{---}} & \mathbf{y}^T_m & \textbf{\text{---}}\\
                    \textbf{\text{---}} & \mathbf{y}^T_m & \textbf{\text{---}}\\
                     & \vdots & \\
                    \textbf{\text{---}} & \mathbf{y}^T_m & \textbf{\text{---}}
    \end{bmatrix}.
\end{equation} 

\section{Computational Complexities}
\label{sec:comp_complex}

In order to analyze the computational complexity of the proposed PT-RBF joint channel estimation and equalization, we rely on the PT-RBF computational complexity presented in~\cite{MayerThesis}~(see Table 3.7 on page 87). In this work, the computational complexities are expressed in terms of the number of real-valued multiplications and additions. To accomplish this, each complex-valued multiplication is represented by four real-valued multiplications and two additions, and each complex-valued addition is represented by two real-valued additions. Based on~\cite{MayerThesis}, Table~\ref{tab:ptrbf_comp_complex} depicts the computational complexities of training and inference of the PT-RBF employed in this work.
\begin{table}[htb]
\caption{\label{tab:ptrbf_comp_complex}Computational complexities for training and inference of the PT-RBF for joint channel estimation and equalization.}
\centering
{\tt
\begin{tabular}{|c||c|c|}\hline
Stage&Additions&Multiplications\\\hline\hline
Training&$N_n(16K-2)+4K$&$N_n(16K+14)+2K$\\\hline
Inference&$N_n(8K-2)$&$N_n(6K+2)$\\\hline
\end{tabular}
}
\end{table}

Then, keeping the number of subcarriers $K=64$, Table~\ref{tab:ptrbf_comp_complex_number} presents the PT-RBF computational complexities depending on the number of neurons $N_n$. Additions are described by $+$, and multiplications by $\times$.
\begin{table}[htb]
\caption{\label{tab:ptrbf_comp_complex_number}PT-RBF computational complexities for training and inference, keeping $K=64$ subcarriers.}
\centering
{\tt
\begin{tabular}{|c||r|r|r|r|}\hline
\multirow{2}{*}{$N_n$}&\multicolumn{2}{|c|}{Training}&\multicolumn{2}{|c|}{Inference}\\\cline{2-5}
&\multicolumn{1}{|c|}{$+$}&\multicolumn{1}{|c|}{$\times$}&\multicolumn{1}{|c|}{$+$}&\multicolumn{1}{|c|}{$\times$}\\\hline\hline
256&261,888&265,856&130,560&98,816\\\hline
512&523,520&531,584&261,120&197,632\\\hline
1,024&1,046,784&1,063,040&522,240&395,264\\\hline
2,048&2,093,312&2,125,952&1,044,480&790,528\\\hline
\end{tabular}
}
\end{table}

For the sake of simplicity, in order to compare the computational complexity of the proposed work with the joint channel estimation and decoding proposed by Ye et al.~\cite{Ye2018}, we only consider the number of real-valued multiplications since it is the most onerous mathematical operation. The number of real-valued multiplications of an RVNN with multiple fully-connected layers can be found in ~\cite{MayerThesis}~(see Table 2.1 on page 37). Unlike the work of Ye et al.~\cite{Ye2018}, the PT-RBF here presented does not implement decoding. Thus, for a fair comparison, we add the FFT computation complexity to the PT-RBF one. The number of real-valued multiplications of the FFT was obtained from~\cite{Sorensen1986}~(see Table~I on page 154, three-BF and length 64 results in $248$ real-valued multiplications). Table~\ref{tab:complexities} presents the computational complexities of the PT-RBF with FFT and the DNNs proposed by Ye et al.~\cite{Ye2018} (8 RVNNs with five layers each, containing 256, 500, 250, 120, and 16 neurons, respectively). Note that, as Ye et al.~\cite{Ye2018} handle two OFDM symbols at a time, we multiply the PT-RBF with FFT computational complexity by two. The number of subcarriers is $K=64$ and the PT-RBF is considered with $N_n=2,048$ neurons (the maximum number of neurons implemented in this work).
\begin{table}[htb]
\caption{\label{tab:complexities}Computational complexities of the PT-RBF with FFT and the work of Ye et al.~\cite{Ye2018}.}
\centering
{\tt
\begin{tabular}{|c||c|c|}\hline
Stage&PT-RBF with FFT&Ye et al.~\cite{Ye2018}\\\hline\hline
Training&4,252,400&5,835,344\\\hline
Inference&1,581,552&2,279,360\\\hline
\end{tabular}
}
\end{table}

From Table~\ref{tab:complexities}, in an equivalent scenario, the proposed approach has approximately $30\%$ less complexity compared with the work of Ye et al.~\cite{Ye2018}. For a similar complexity, we could extend the PT-RBF to about $N_n=2,900$ neurons. Moreover, it is important to highlight that in the inference phase, our approach generates $100\%$ of useful information since no pilot is necessary. On the other hand, the useful information rate (not taking the CP into account) of Ye et al.~\cite{Ye2018} is given by $100\times (N_p-K)/N_p$ $[\%]$, where $N_p$ is the number of pilots per OFDM symbol. Then, for $K=64$ subcarriers and $N_p=8$ pilots, the useful information rate is $87.5\%$. Consequently, the proposed approach presents a lower computational complexity with a higher useful information rate.

\section{Results}
\label{sec:results}

In order to represent a practical scenario, we set the simulation system with the 3GPP TS 38.211 specification for 5G physical channels and modulation~\cite{3gpp.38.211}. The OFDM is defined with 240~kHz subcarrier spacing, 64 active subcarriers, and a block-based pilot scheme with a preamble of $5,000$ OFDM symbols. Symbols are modulated with quadrature phase shift keying~(QPSK).

Based on the tapped delay line-A~(TDL-A) from the 3GPP~TR~38.901 5G channel models~\cite{3gpp.38.901}, the massive MIMO channel follows the TDLA100 from the 3GPP~TR~38.104 5G radio base station transmission and reception~\cite{3gpp.38.104}. The channel is described with 12 taps, with varying delays from 0.0~ns to 290~ns and powers from -26.2~dB to 0~dB. A Rayleigh distribution is used to compute the sample-spaced multipath channel $\mathbf{h}[n]$.

The PT-RBF operates with 64 inputs and outputs and one hidden layer with $N_n$ Gaussian neurons. The inputs are taken from the FFT outputs. The PT-RBF outputs are equalized symbols. Each OFDM symbol from the preamble is handled as a sample for training. The PT-RBF is trained for $200$ epochs with a shuffle, to improve convergence. The learning rates were optimized, by trial and error, as $\eta_w=0.02$, $\eta_b=0.02$, $\eta_\gamma=0.02$, and $\eta_\sigma=0.01$.

\begin{figure}[t]
\centering  \includegraphics[width=\columnwidth]{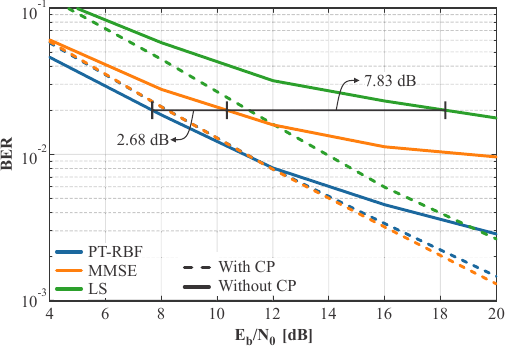}
    \caption{BER results of the proposed approach, MMSE, and LS channel estimation algorithms with and without CP. Solid lines regard results without CP and dashed lines with CP.}
    \label{fig:result_estimators}
\end{figure}

Fig.~\ref{fig:result_estimators} shows the results of the proposed PT-RBF for joint channel estimation and the classical MMSE and LS algorithms. In this comparison, we have set $N_n=2048$ neurons. It is important to highlight that, as results are discussed in terms of bit energy to noise power spectral density ratio ($\mathrm{E_b}/\mathrm{N}_0$), we take the CP into account to plot the bit error rate~(BER). Also, we consider a pre-FEC~(pre-forward error correction) BER of $2\times10^{-2}$~\cite{Castro2019} for comparison. In the best-case scenario, with a CP long enough to cover all channel impulses~(17~samples), the PT-RBF presented similar results with MMSE, and the LS achieved an inferior performance of 2.91~dB. On the other hand, in the worst-case scenario (i.e., without CP), the PT-RBF approach presents better performance, surpassing the MMSE and LS by about 2.68~dB and 10.51~dB, respectively. This better performance follows the results demonstrated in~\cite{Ye2018}, but with lower computational complexity.

\begin{figure}[t]
\centering
\includegraphics[width=\columnwidth]{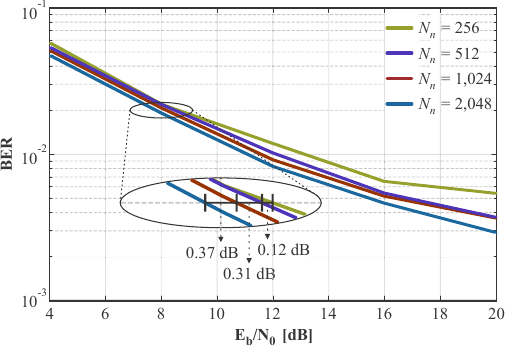}
    \caption{PT-RBF joint channel estimation and equalization results depending on the number of neurons $N_n$. Results magnified for a BER $=2\times10^{-2}$ to highlight performance.}
    \label{fig:result_neurons}
\end{figure}

Fig.~\ref{fig:result_neurons} shows the PT-RBF performance without CP, depending on the number of neurons $N_n$. In Fig.~\ref{fig:result_neurons}, we magnify the $\mathrm{E_b}/\mathrm{N}_0$ results for a BER $=2\times10^{-2}$ for comparison. As expected, the performance decays with $N_n$ reduction. However, reducing the number of neurons by half, from $2,048$ to $1,024$, only impacts $0.37$~dB performance. When reducing the complexity by one eighth (i.e., from $2,048$ to $256$ neurons), the $\mathrm{E_b}/\mathrm{N}_0$ is worsened by only $0.8$~dB.

\section{Conclusions}
\label{sec:conc}


This work proposes a complex-valued neural network (CVNN) for joint channel estimation and equalization in OFDM systems without a cyclic prefix. Unlike previous works in the literature, where neural networks were employed to handle both channel estimation and decoding, we focus solely on addressing channel imperfections. As a result, our neural network does not need to learn the FFT decoding, thereby reducing its computational complexity. It is preferable to keep the FFT mapping outside of the neural network since it can be implemented with very low computational complexity. Our results demonstrate that the proposed PT-RBF outperforms both MMSE and LS algorithms. Furthermore, the computational complexity can be reduced by half at the cost of an additional~0.37~dB. In future works, we plan to implement other well-known CVNNs and to test for other system and channel imperfections, such as channel dynamics and nonlinearities.

\bibliographystyle{IEEEtran}
\bibliography{bibliography.bib}

\end{document}